# MARKOV CHAIN MONTE CARLO WITH GAUSSIAN PROCESS EMULATION FOR A 1D HEMODYNAMICS MODEL OF CTEPH

**Amirreza Kachabi, Mitchel J. Colebank, Sofia Altieri Correa, Naomi C. Chesler**
Edwards Lifesciences Foundation Cardiovascular Innovation and Research Center, Department of Biomedical Engineering, University of California, Irvine, Irvine, CA, USA
`{akachabi,mcoleban,altieris,nchesler}@uci.edu`

## SUMMARY

Microvascular disease is a contributor to persistent pulmonary hypertension in those with chronic thromboembolic pulmonary hypertension (CTEPH). The heterogenous nature of the micro and macrovascular defects motivates the use of personalized computational models, which can predict flow dynamics within multiple generations of the arterial tree and into the microvasculature. Our study uses computational hemodynamics models and Gaussian processes for rapid, subject-specific calibration using retrospective data from a large animal model of CTEPH. Our subject-specific predictions shed light on microvascular dysfunction and arterial wall shear stress changes in CTEPH.

## 1 INTRODUCTION

Chronic thromboembolic pulmonary hypertension (CTEPH) is a subgroup of pulmonary hypertension (PH) caused by blood clots that lodge in the blood vessels of the lungs. Even though CTEPH can be treated with surgery, persistent PH affects some patients, especially those who suffer from microvascular disease, which cannot be addressed surgically. Persistent PH has been identified as the primary cause of mortality in CTEPH, responsible for nearly 75% of early postoperative deaths [1]. Quantification of micro and macrovascular disease and severity can provide important insight into the development of persistent PH and prognosis after surgery [2].

Computational hemodynamics models and simulations have become useful tools in the medical field, providing important information that are infeasible to measure clinically. However, the parameters in these models need to be specified and calibrated based on each subject individually. Hence, one aspect of a useful and practical computational model is how quickly and accurately it can be calibrated for each subject. Identifying the uncertainties in parameters and outputs are also crucial.

In this study, we investigate changes in the structure and function of the arterial micro and macrovasculature using a previously developed one-dimensional (1D) fluid dynamics model (a system of partial differential equations, PDE) of a large animal model of CTEPH. We use Gaussian processes (GPs) to emulate the main mathematical model to enable calibration to the subject data quickly and accurately. This approach reduces the computation time from a month to less than a day, making it appealing in the medical industry. Finally, we investigate the changes in wall shear stress (WSS), which is thought to correlate with endothelial cell dysfunction [3]. We examine how CTEPH impacts arterial WSS predictions in the micro and macro pulmonary vasculature of both lungs.

## 2 METHODOLOGY
### 2.1 Animal study and data collection
All experimental data were acquired retrospectively from [4]. Data used from the animal study are magnetic resonance (MR) angiography imaging data, systolic and diastolic main pulmonary artery (MPA) pressure ($p_{sys}^{MPA}, p_{dia}^{MPA}$), time series flow at the main, left and right pulmonary arteries ($\boldsymbol{q}^{MPA}, \boldsymbol{q}^{LPA}, \boldsymbol{q}^{RPA}$) and time series MPA area ($\boldsymbol{A}^{MPA}$) in both pre (baseline) and post CTEPH stages. We create a 1D network from MR data as done previously [5]. The calibration data are $\boldsymbol{y} = \{p_{sys}^{MPA}, p_{dia}^{MPA}, \boldsymbol{q}^{LPA}, \boldsymbol{q}^{RPA}, \boldsymbol{A}^{MPA}\}$.

## 2.2 Fluid mechanics and model descriptions

We use time-dependent, 1D fluid dynamics equations to simulate pulmonary arterial macrovascular hemodynamics, as described previously [5]. We employ a linear elastic stress-strain relationship assuming a thin-walled, isotropic, incompressible cylinder. The constitutive law is

$$P(x,t) = \frac{4}{3} K \left( \sqrt{A(x,t)/A_{dias}} - 1 \right) + P_0, \tag{1}$$

where $K$ (g/cm s²) accounts for structural and load-dependent arterial stiffness, and $A_{dia}$ (cm²) is the diastolic area obtained during imaging. The term $P_0 = 8$ (mmHg) denotes the pulmonary capillary wedge pressure (PCWP), which represents the left atrial pressure. Here we use PCWP in place of diastolic pressure as it is a better representation of the reference pressure in the pulmonary circuit. The inflow boundary condition is the measured time series flow data in the MPA. We used structured tree outlet boundary conditions, where each terminal vessel within the network of large arteries is connected to a network of synthetic vessels that represents the microvasculature [6]. The structured tree is described by the area ratio, $\zeta$, and the radius exponent $\gamma$, using the relationships

$$\zeta = \frac{A_{d2}}{A_{d1}}, \quad r_p^\gamma = r_{d1}^\gamma + r_{d1}^\gamma, \gamma \in [2.33, 3]. \tag{2}$$

The daughter areas, $A_{d2}$ and $A_{d1}$, correspond to the smaller and larger daughter vessels in each bifurcation respectively, and $r_p$ is the parent radius and $r_1$ and $r_2$ are its daughter radii. $\zeta$ was calculated as the median among all bifurcations across the large arterial tree, with $\zeta = 0.6$. The length of the vessels in the structured tree are calculated by a length-to-radius ratio, $lrr$. The structured tree terminates when the vessel radii are $< r_{min} = 0.005$ cm, i.e. less than 50 microns [6].

The WSS is computed as $\tau(x,t) = \mu(\partial u/\partial r)_{r=R}$, with

$$u(x,t) = \bar{U}(t) \frac{\eta + 2}{\eta} \left( 1 - \frac{r^\eta}{R(x,t)^\eta} \right), \tag{3}$$

where $\mu$ is the dynamic blood viscosity (g/cm), $u(x,t)$ is the blood velocity (cm/s), $\bar{U}(t)$ is the mean velocity (cm/s), $R(x,t)$ is the outer radius of the vessel (cm) and $\eta$ is the power-law constant that is set to be 5 based on our previous study [5].

## 2.3 Parameter values

There are two different sets of parameters. Constant parameters are fixed based on measured data and literature, such as heart rate, blood viscosity, blood density, PCWP and $\zeta$. A different set of parameters are inferred, including vessel stiffness ($K$), radius relationship ($\gamma$), and length to radius ratio ($lrr$). The latter two parameters modulate vascular impedance [6]. Since CTEPH is a heterogeneous disease that can affect each side of the lung differently, we used independent $\zeta$ and $lrr$ parameters for the left and right lungs. In total, our inverse problem is a five-dimensional problem, comprised of $\boldsymbol{\theta} = \{K, \gamma_L, \gamma_R, lrr_L, lrr_R\}$.

## 2.4 Emulation

The computational cost of the PDE simulator is 15 minutes; thus, we constructed a GP emulator to speed up computation. We trained the GP model, $m(\boldsymbol{\theta})$, on 2,000 datasets from the PDE simulator with different combinations of $\boldsymbol{\theta}$ using a Latin hypercube design. The outputs consist of four different time-varying signals: MPA pressure, LPA and RPA flows, and MPA area, making the final output dimension 32*4 = 128. We employed min-max scaling for both the inputs and outputs and applied principal component analysis (PCA) to the outputs. We reduced the dimension from 128 to 12 principal components, which captures more than 97% of the original variance. Since the principal components of the outputs are independent, we emulate the PCA data using 12 independent GPs [7] with a Matérn covariance kernel ($\nu = 5/2$).

## 2.5 Inverse problem

We use the delayed rejection adaptive Metropolis (DRAM) algorithm using the `pymcmcstat` package [8]. We use uniform prior distributions for the parameters and run DRAM for 12,000 iterations, using the last 2000 samples of the chain for analysis. We check for convergence using Geweke's diagnostic. Our likelihood function for the four outputs is

$$\mathcal{L}(\mathbf{y}|\boldsymbol{\theta}) = \frac{1}{(2\pi)^{N_t/2}} \det(\boldsymbol{\Sigma}_y)^{-1/2} \exp\left(-\frac{1}{2}(\mathbf{y} - \widetilde{\boldsymbol{m}}(\boldsymbol{\theta}))^\top \boldsymbol{\Sigma}_y^{-1}(\mathbf{y} - \widetilde{\boldsymbol{m}}(\boldsymbol{\theta}))\right). \tag{4}$$

The likelihood includes $N_t = 98$ data points corresponding to the data, $\mathbf{y}$, $\widetilde{\boldsymbol{m}}(\boldsymbol{\theta})$ is the inverse PCA transformed GP model predictions, and the measurement covariance $\boldsymbol{\Sigma}_y$, contains the homoscedastic measurement error for each data source along the diagonal. Each measurement error is assumed to be independent, and is updated according to an inverse-gamma distribution [8].

## 3 RESULTS AND CONCLUSIONS

Figure 1 shows changes in the marginal posterior densities in one subject at baseline and CTEPH. CTEPH increases $K$ relative to baseline. This can be attributed to a rise in pulmonary vascular resistance, resulting in increased arterial pressure, fibrotic changes in the pulmonary arterial wall, and structural remodeling. Both $\gamma_L, \gamma_R$ decrease in CTEPH, suggesting microvascular area reductions due to emboli. This finding is consistent with results presented by Olufsen et al. [6]. Alternatively, both $lrr_L$ and $lrr_R$ increase, which elevates pulmonary vascular resistance and pulmonary vascular impedance.

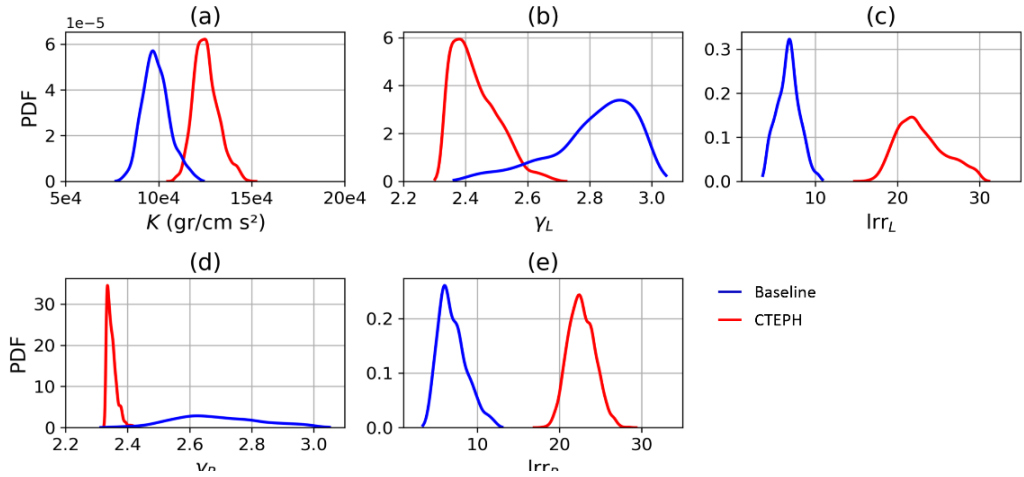

Figure 1: Changes in model parameter posteriors at baseline (blue) and CTEPH (red).

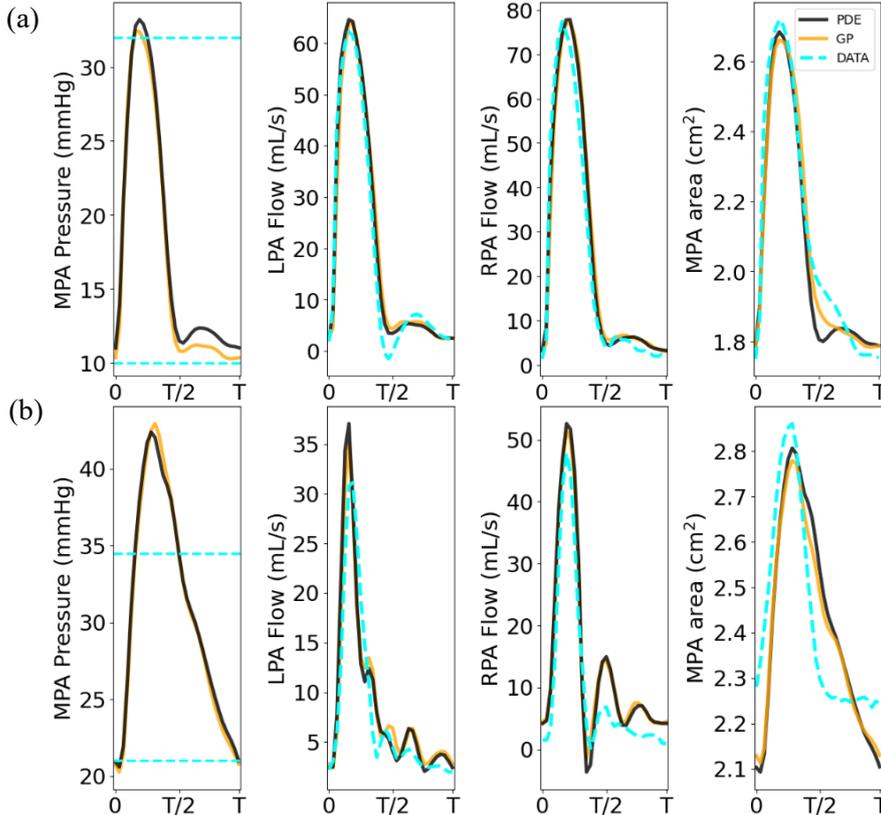

Figure 2: Measured data (cyan) compared to the PDE solutions (black) and the GP emulator (orange).

Figure 2 shows the agreement between the measured data, PDE simulations and the GP predictions at both baseline (a) and CTEPH (b). Results show that both baseline and CTEPH GPs are accurate emulators of the PDE. At baseline, the predictions align with all measured data. Similarly, CTEPH flow predictions accurately capture the measured signal. However, there is a mismatch between the data and predictions of systolic pressure and diastolic area. These discrepancies may be attributed to the limited data points available for pressure calibration. Area predictions are not as accurate in CTEPH, possibly due to our use of baseline MR angiography data for the 1D network.

Figure 3 shows the changes in WSS from baseline to CTEPH in one selected branch on both sides of the lungs. The posterior distributions are used to compute credible intervals for the model response. CTEPH dramatically decreased WSS in the left and right lung. Decreased WSS will drive endothelial cell signaling, leading to vasoconstriction and the production of pro-inflammatory factors as well as collagen synthesis and accumulation [3]. Factors released from endothelial cells in response to decreased WSS may also drive smooth muscle cell (SMC) proliferation [3]. Collagen synthesis and accumulation and SMC proliferation are consistent with the increased $K$ relative to baseline.

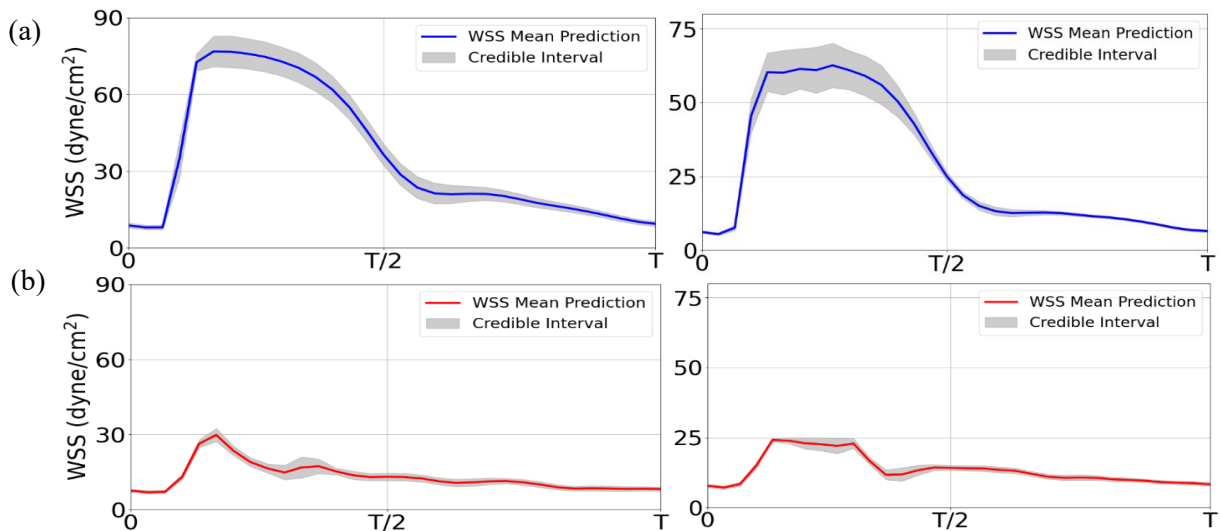

Figure 3: WSS changes from baseline (a) to CTEPH (b) in in a left (left) and right (right) lung branches.

Through the use of advanced, subject-specific modeling techniques and a relatively inexpensive statistical model, we conducted efficient model calibration. A more comprehensive understanding of CTEPH can facilitate the development of targeted interventions tailored to the specific needs of affected individuals, particularly when the microvasculature is severely affected. Further analyses will use the same approach in four additional subjects, both pre and post CTEPH. This expanded investigation will provide a more comprehensive understanding of microvascular contributions to CTEPH and post-surgical prognoses.